\documentclass[aps,prd,twocolumn,superscriptaddress,preprintnumbers,floatfix,nofootinbib,notitlepage,showkeys,showpacs]{revtex4-1}

\usepackage[utf8]{inputenc}

\usepackage{graphicx}
\usepackage{hyperref}
\usepackage{latexsym}
\usepackage{amsmath}
\usepackage{amssymb}
\usepackage{bbm}
\usepackage{ulem}
\usepackage{pdfsync}
\usepackage{epsfig}
\usepackage{epstopdf}
\usepackage{subfigure}
\usepackage[dvipsnames]{xcolor}
\usepackage{comment}
\usepackage{slashed}
\usepackage{multirow}

\begin{document}

\title{The trouble beyond $H_0$ and the new cosmic triangles}
\author{Jos\'e Luis Bernal}
\affiliation{Department of Physics and Astronomy, Johns Hopkins University, 3400 North Charles Street, Baltimore, Maryland 21218, USA}
\author{Licia Verde}
\affiliation{ICC, University of Barcelona, Mart\'i  i Franqu\`es, 1, E-08028 Barcelona, Spain}
\affiliation{ICREA, Pg. Lluis Companys 23, Barcelona, 08010, Spain.} 
\author{Raul Jimenez}
\affiliation{ICC, University of Barcelona, Mart\'i  i Franqu\`es, 1, E-08028 Barcelona, Spain}
\affiliation{ICREA, Pg. Lluis Companys 23, Barcelona, 08010, Spain.} 
\author{Marc Kamionkowski}
\affiliation{Department of Physics and Astronomy, Johns Hopkins University, 3400 North Charles Street, Baltimore, Maryland 21218, USA}
\author{David Valcin}
\affiliation{ICC, University of Barcelona, Mart\'i  i Franqu\`es, 1, E-08028 Barcelona, Spain}
\author{Benjamin D. Wandelt}
\affiliation{Sorbonne Universit\'e, CNRS, UMR 7095, Institut d'Astrophysique de Paris, 98 bis bd Arago, 75014 Paris, France.}
\affiliation{Sorbonne Universit\'e, Institut Lagrange de Paris (ILP), 98 bis bd Arago, 75014 Paris, France.}
\affiliation{Center for Computational Astrophysics, Flatiron Institute, 162 5th Avenue, 10010, New York, NY, USA.}

\begin{abstract}
The distance ladder using supernovae yields higher values of the Hubble constant $H_0$ than those inferred from measurements of the cosmic microwave background (CMB) and galaxy surveys, a discrepancy that has come to be known as the `Hubble tension'. This has motivated the exploration of extensions to the standard cosmological model in which higher values of $H_0$ can be obtained from CMB measurements and galaxy surveys. The trouble, however, goes beyond $H_0$; such modifications affect other quantities, too. In particular, their  effects on cosmic times are usually neglected. We explore here the implications that measurements of the age $t_{\rm U}$ of the Universe, such as a recent inference from the age of the oldest globular clusters,  can have for potential solutions to the $H_0$ tension. The value of $H_0$ inferred from the CMB and galaxy surveys is related to the sound horizon at CMB decoupling (or at radiation drag), but it is also related to the matter density and to $t_{\rm U}$.  Given this observation, we show how model-independent measurements may support or disfavor proposed new-physics solutions to the Hubble tension. Finally, we argue that cosmological measurements today provide constraints that, within a given cosmological model, represent an over-constrained system, offering a powerful diagnostic tool of consistency. We propose the use of ternary plots to simultaneously visualize independent constraints on key quantities related to $H_0$ like $t_{\rm U}$, the sound horizon at radiation drag, and the matter density parameter. We envision that this representation will help find a solution to the 
trouble of  and beyond $H_0$.
\end{abstract}

\maketitle
\section{Introduction}
The standard, $\Lambda$CDM, cosmological model, has successfully passed increased scrutiny, as observations of the cosmic microwave background (CMB)~\cite{Planck18_pars,SPTpol,ACT_2020}, type-Ia supernovae (SNeIa)~\cite{Scolnic_pantheon} and large-scale structure~\cite{Alam_bossdr12, DESY1_cosmo, Heymans_kids1000cosmo,eBOSS_cosmo} have improved drastically over recent years. Nonetheless, tensions have arisen for specific parameters when their values are  inferred,  within the $\Lambda$CDM, from different probes and observables. The biggest tension is related to determinations of  the Hubble constant $H_0\equiv 100h$ km/s/Mpc, and has increased in the last decade to be in the $4-5\sigma$~\cite{Verde_tension, Verde_KITP}. 

\begin{figure}[t]
 \begin{centering}
\includegraphics[width=\columnwidth]{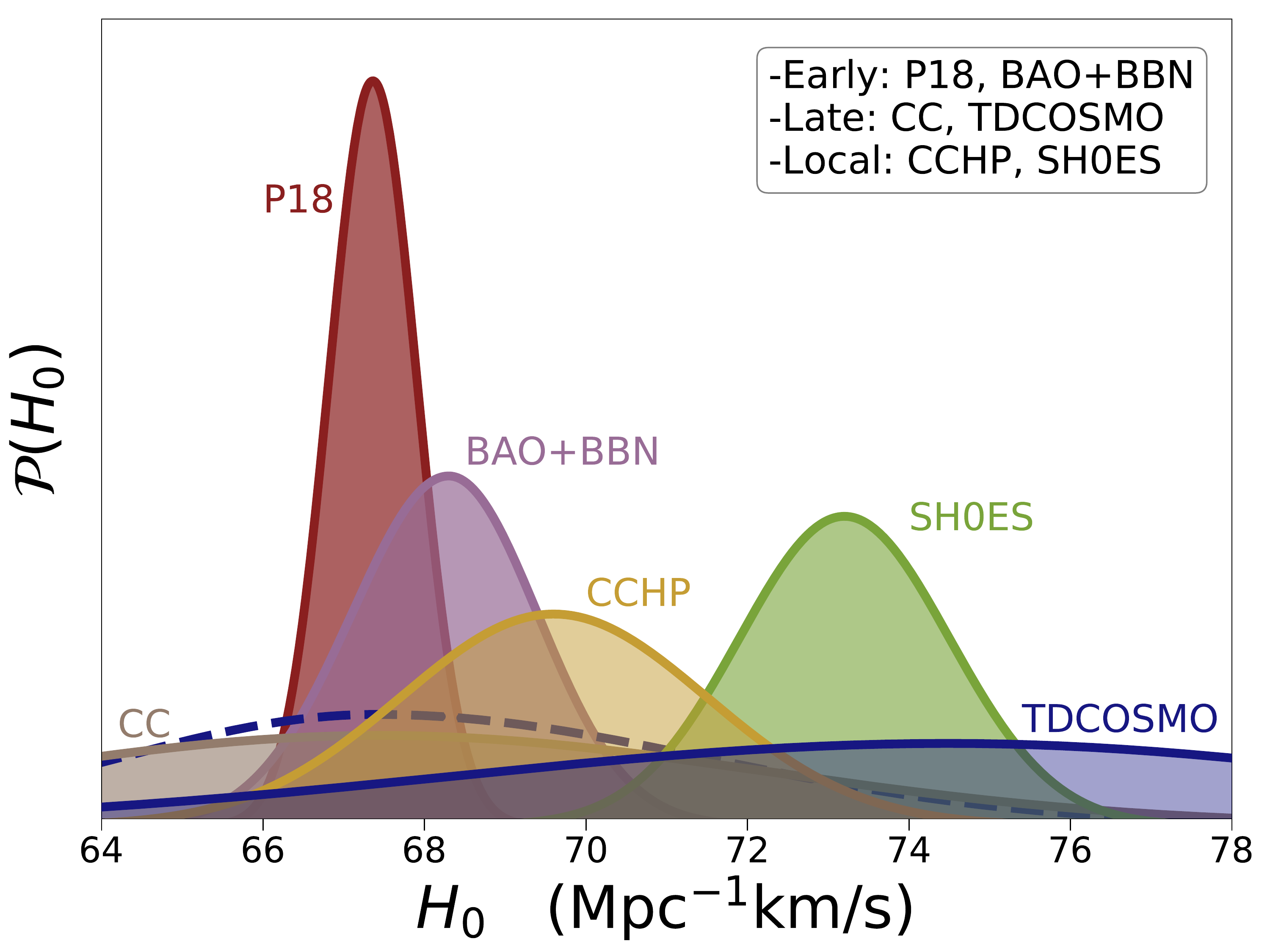}
\caption{Summary of constraints on $H_0$ from cosmic chronometers (CC)~\cite{Haridasu_clocks}, \textit{Planck}  (P18)~\cite{Planck18_pars}, baryon acoustic oscillations with a BBN prior on the baryon abundance (BAO+BBN)~\cite{Schoneberg_BAOBBN}, CCHP~\cite{Freedman_H020}, SH0ES~\cite{Riess_H020}, and strong-lensing time delays (TDCOSMO)~\cite{TDCOSMO_IV}. We also show (dashed line) the TDCOSMO constraint including resolved kinematics from SLACS galaxies, which assumes both samples belong to the same parent population. Note that the results shown in this figure are subject to different model assumptions.}  
\label{fig:H0}
\end{centering}
\end{figure}

The current state of the $H_0$ tension is illustrated in Fig.~\ref{fig:H0}, where we show marginalized posteriors  for  measurements depending on early-times physics (like \textit{Planck}~\cite{Planck18_pars}  or baryon acoustic oscillations with a big bang nucleosynthesis prior on the physical density of baryons~\cite{Addison_BAOBBM, Schoneberg_BAOBBN}), late-time expansion history (using strong lensing time delays from TDCOSMO~\cite{Chen_sharp, Wong_holicow2020, Shajib_strides, Millon_TDCOSMO, TDCOSMO_IV}\footnote{There are ongoing efforts to relax the dependence of strong lensing time delays $H_0$ inference on the assumed expansion rate~\cite{Chen_TDCOSMOVI}.} and cosmic chronometers~\cite{Jimenez_C, Haridasu_clocks}), and local measurements, independent of cosmology, from SH0ES ~\cite{RiessH0_19} and CCHP \cite{Freedman_H020}. Except for cosmic chronometers, all competitive $H_0$ constraints considered here rely on distance measurements.\footnote{Some $H_0$ constraints related with large-scale structure do not depend on  the sound horizon, but still depend on distance scales, such as the size of the horizon at matter-radiation equality~\cite{Baxter_H0CMBlensing, Philcox_H0nord}.}

The two determinations yielding  the largest tension are obtained from the CMB power spectra and the SH0ES
distance ladders using SNeIa calibrated by Cepheids. CCHP calibrates the SNeIa instead  with the tip of the red giant branch (TRGB) and  finds a lower value of $H_0$~\cite{Freedman_H020} (see also~\cite{Yuan_TRGBH0, Reid_TRGBNGC4258,Nataf_LMCred}). 

Given the strong constraints imposed by available data on the product of the sound horizon $r_{\rm d}$ at radiation drag and $h$, $r_{\rm d}$ has been targeted as the critical quantity to be modified in order to solve the $H_0$ tension. Baryon acoustic oscillations (BAO) and SNeIa disfavor any strong deviation from the evolution of the expansion rate predicted $\Lambda$CDM, while strongly constraining  $r_{\rm d}h$~\cite{BernalH0,Poulin_H0,Aylor_sounds}. In light of current constraints, the modifications of $\Lambda$CDM best poised to reduce the $H_0$ tension involve altering pre-recombination physics as to lower the value of $r_{\rm d}$, as it is discussed in Ref.~\cite{Knox_H0hunter}, where it is argued that other possibilities, both before and after recombination are disfavored by observations or theoretically unlikely. There is a plethora of proposed  models to do so and  those showing more promise involve boosts of the expansion history between matter-radiation equality and recombination (see e.g.,~\cite{Karwal_EDE, Poulin_EDE, Smith_EDE, Lin_acousticDM, Agrawal_R&R, Berghaus_H0friction, Sakstein_H0nu, Niedermann_nEDE, Zumalacarregui_MGH0, Braglia_H0evolG, Ballardini_H0STgrav, Braglia_alphaEDE,Ballesteros_MGH0, Niedermann_nEDE20, Braglia_H0earlyMG, Abadi_MGH0}). 

Despite the fact that  most of the attention has been focused on modifying distance scales across cosmic history,
the expansion rate, thus $H_0$, also  determines the age-redshift relation. Measuring cosmic ages can provide a constraint on $H_0$ completely independent from $r_{\rm d}$, other standard scales, or distance measurements (see e.g.,~\cite{Jaffe_H0t} for a study regarding the presence of a cosmological constant).
Cosmic chronometers measure directly the expansion rate using differential ages~\cite{Jimenez_C}; this approach is  limited to relatively low redshifts, covering a range that overlaps with  distance measurements. 
On the other hand, since relative changes in the expansion history at early times do not significantly modify the age of the Universe, independent inferences of absolute lookback times, such as the age of the Universe,  may weigh in on the $H_0$ tension. 

In this work, we discuss how the age of the Universe inferred from a recent determination of the age of the oldest globular clusters~\cite{Jimenez_GC,Valcin_GC, Valcin_GCsyst} can offer an additional perspective on the $H_0$ controversy.  Our results suggest that an accurate and precise measurement of the age of the Universe provides an important test of the hypothesis that the $H_0$ tension suggests new early-Universe physics but standard late-Universe physics.  in the process, we also update constraints on the low-redshift expansion rate using recent relative distance redshift measurements.  

In the same way as the $H_0$ tension was reframed as the inconsistency between $r_{\rm d}$, $h$ and  their product $r_{\rm d}h$ (inferred independently in a model-agnostic way from low redshifts observations)~\cite{BernalH0,Poulin_H0,Aylor_sounds},  the same can be said about other sets of quantities that can be constrained independently, albeit assuming a cosmological model. One is the combination of  the matter density parameter $\Omega_{\rm M}$ today, $h^2$, and  their product, the physical matter density $ \Omega_{\rm M}h^2$. 
The other  set is the age $t_{\rm U}$ of the Universe and $h$, and their combination $t_{\rm U}h$, which is completely determined by the shape of the expansion history and measured independently. 

This is reminiscent of the `cosmic triangle'  proposed in Ref.~\cite{Bahcall_triangle} two decades ago, where the matter, cosmological constant and curvature density parameters  are related to one another because they sum to unity.  The original cosmic triangle is a ternary plot which served  to visualize cosmological constraints that led to favor the (now standard) flat $\Lambda$CDM model. 
Here, in full analogy, we propose the use of ternary plots as diagnosis diagrams to examine the tension between cosmological quantities independently measured from different observations. Ternary plots are specially suited for this purpose, as we show for the cases of $r_{\rm d}$, $\Omega_{\rm M}$ and $t_{\rm U}$ listed above.

This article is organized as follows. We present updated constraints on the late-Universe expansion rate as a function of redshift in Sec.~\ref{sec:data_and_Ez}; discuss the role cosmic ages play in the $H_0$ tension in Sec.~\ref{sec:tU_tension}; present the new cosmic triangles in Sec.~\ref{sec:triangles}; and finally conclude in Sec.~\ref{sec:conclusions}.

\section{Updated expansion rate constraints}
\label{sec:data_and_Ez}
We begin by presenting updated model-agnostic constraints on the expansion rate  as a function of redshift, $E(z)\equiv H(z)/H_0$, using the latest, state-of-the-art data.
These constraints on  $E(z)$ are a key input for  the results of sections \ref{sec:tU_tension}, \ref{sec:triangles} and our conclusions.

We use SNeIa observations from Pantheon~\cite{Scolnic_pantheon} and BAO measurements from 6dFGRS~\cite{Beutler11}, SDSS DR7~\cite{Ross15}, BOSS~\cite{Alam_bossdr12}, and eBOSS, including galaxies, quasars and Lyman-$\alpha$ forest~\cite{GilMarin_eBOSSDR16_LRG,Raichoor_eBOSS_BAOELG,Hou_eBOSSDR16_QSO,Neveux_eBOSSDR16_QSO,duMasdesBourboux_eBOSS16_LyaF} as relative distance indicators.\footnote{Standard BAO analyses adopt a prior on $r_{\rm d}$ to break the $r_{\rm d}h$ degeneracy and calibrate the distance measurements, following the approach known as inverse cosmic distance ladder. Not using that prior and marginalizing over $r_{\rm d}$ removes any dependence on pre-recombination physics, since the BAO measurements are robust to modifications of the pre-recombination physics of $\Lambda$CDM~\cite{Bernal_BAObias}. 

We use measurements from BAO-only analyses, following the eBOSS likelihoods and criterion to combine with BOSS measurements from \url{https://svn.sdss.org/public/data/eboss/DR16cosmo/tags/v1_0_0/likelihoods/BAO-only/}.} Note that, although BAO-only analyses assume a fiducial cosmology, their results are robust to be applied to other cosmologies (see e.g.,~\cite{Carter_BAOtest,Bernal_BAObias}).

Two models for $E(z)$ are examined: $\Lambda$CDM, and a parametrization using natural cubic splines, the nodes of which have a varying position, without imposing flatness, which we refer to as `generic' expansion and as such falls under what we here refer to as ``model agnostic" approach. Given its flexibility, the generic expansion shall be understood as a marginalization over cosmological models predicting a smooth $E(z)$.  Other uses of this parametrization, known as flexknot, can be found in e.g., Refs.~\cite{Vazquez_flexknot, Millea_flexknot}. 

 The free parameters for the $\Lambda$CDM case are $\left\lbrace \Omega_{\rm M},\, r_{\rm d},\,h,\, M_{\rm SN}\right\rbrace$, where $M_{\rm SN}$ is the absolute magnitude of SNeIa; on the other hand, the generic expansion needs $\left\lbrace \boldsymbol{z}_{\rm knot}^{(1,N-1)},\,\boldsymbol{E}_{\rm knot}^{(1,N)},\, \Omega_{\rm k},\, r_{\rm d},\,h,\, M_{\rm SN} \right\rbrace$ as free parameters, where $\boldsymbol{E}_{\rm knot}$ are the values of $E(z)$ at the knots of the splines, located at $\boldsymbol{z}_{\rm knot}$, and $\Omega_{\rm k}$ is the density parameter associated to curvature. The first and last knot are fixed at $z=0$ and $z=2.4$, respectively, and $E(0)=1$ by definition. Although our results do not significantly depend on the number of knots used, we find $N=4$ provides the best performance, allowing for as much freedom as possible but avoiding over-fitting and  ${\rm d}E/{\rm d}z$ changing sign too many times, and report the results obtained under this choice. We use uniform priors in all cases.

We use the public code \texttt{MABEL}\footnote{\url{https://github.com/jl-bernal/MABEL}}~\cite{Chen_SlensEz}, to run run Monte Carlo Markov chains with the sampler \texttt{zeus}~\cite{zeus,ess}\footnote{\url{https://zeus-mcmc.readthedocs.io/}} to constrain the shape of the expansion rate in the late-time Universe ($z\leq 2.4$) and the quantity $r_{\rm d}h$ with uncalibrated distance measurements from BAO and SNeIa measurements. Note that, with the data included in the analysis, $h$ and $r_{\rm d}$ individually  are  completely unconstrained; only their product is constrained.

\begin{figure}[t]
\begin{centering}
\includegraphics[width=\columnwidth]{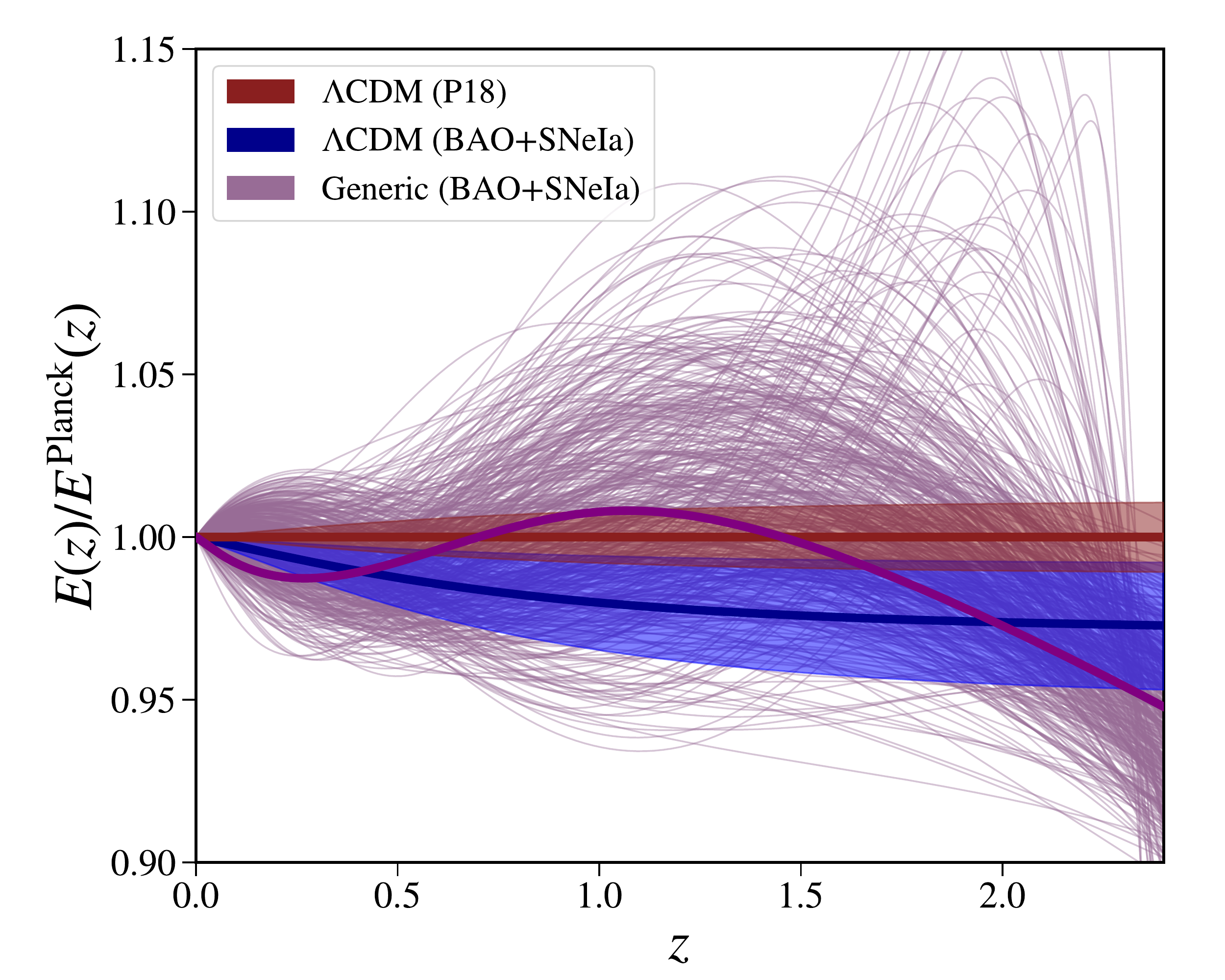}
\caption{Best fit evolution of the expansion rate with redshift (thick lines) normalized by \textit{Planck}'s $\Lambda$CDM best fit ($E(z)/E^{\rm Planck}(z)$) and 68\% confidence level uncertainties (shaded regions, thin lines). \textit{Planck}'s $\Lambda$CDM results are reported in red and  BAO+SNeIa constraints assuming $\Lambda$CDM are in blue. In purple, the reconstruction from BAO+SNeIa assuming a generic expansion; thin lines are a sample of 500 flexknot splines reconstruction from the 68\% cases with highest posterior.
}  
 \label{fig:Ez}
\end{centering}
\end{figure}

The new BAO and SNeIa data allow  the constraints on the generic $E(z)$ to be extended up to $z=2.4$, as shown in Fig.~\ref{fig:Ez}. The generic reconstruction yields an  $E(z)$ which is consistent with the prediction of a $\Lambda$CDM model from \textit{Planck} and BAO+SNeIa. Allowed  deviations from  \textit{Planck}'s $\Lambda$CDM best fit are $\lesssim 3-4\%$ at $z\lesssim 0.8$; this bound weakens slightly  $\lesssim 10\%$ at $0.8\lesssim z\lesssim 2.4$, due to the degradation in the constraining power of SNeIa observations. 
While still being consistent with the $\Lambda$CDM prediction, the reconstructed posterior allows for a boost of the expansion rate ($\sim 15\%$ larger than \textit{Planck}'s $\Lambda$CDM best fit) at $1.5\lesssim z \lesssim 2.4$, this can  be seen as an ``excess wiggle" in the plot; however it is not significant and  we should remark that there are no measurements in that redshift range corresponding to  the gap between the redshift covered by Supernovae data/eBOSS quasars and the Lyman-$\alpha$ forest data. Note also that those expansion histories  showing an excess expansion rate at these redshifts need a lower $E(z)$ than $\Lambda$CDM at low redshifts. These results extend and improve previous constraints from agnostic reconstructions of $E(z)$ (see e.g., Ref.~\cite{BernalH0}, where reported 68\% confidence level limits of the deviations are $~5\%$ at $z\lesssim 0.6$  but grow significantly  at higher redshift). 

Moreover, we find $\Omega_{\rm k}=-0.02\pm 0.10$ and $r_{\rm d}h=100.3\pm 1.2$ Mpc which  represent, respectively, a factor of 6 and factor of 2 improvement compared to the results reported in Ref.~\cite{StandardQuantities} (although the parametrization of $E(z)$ is different, so this comparison is more qualitative than strictly  quantitative; the improvement is driven by the new data gathered over the past five years). These constraints can be compared to those obtained also from BAO+SNeIa when assuming a flat $\Lambda$CDM model: $r_{\rm d}h=100.6\pm 1.1$ Mpc and $\Omega_{\rm M}=0.297\pm 0.013$. As can be seen, the generic reconstruction, despite having five extra model parameters, does not degrade the  $\Lambda$CDM $r_{\rm d}h$ constraints. Furthermore, it returns constraints on $r_{\rm d}h$ comparable to \textit{Planck} results assuming $\Lambda$CDM ($r_{\rm d}h=99.1 \pm 0.9$ Mpc), without relying on early-time physics or observations.

\begin{figure*}[t]
 \begin{centering}
\includegraphics[width=0.32\textwidth]{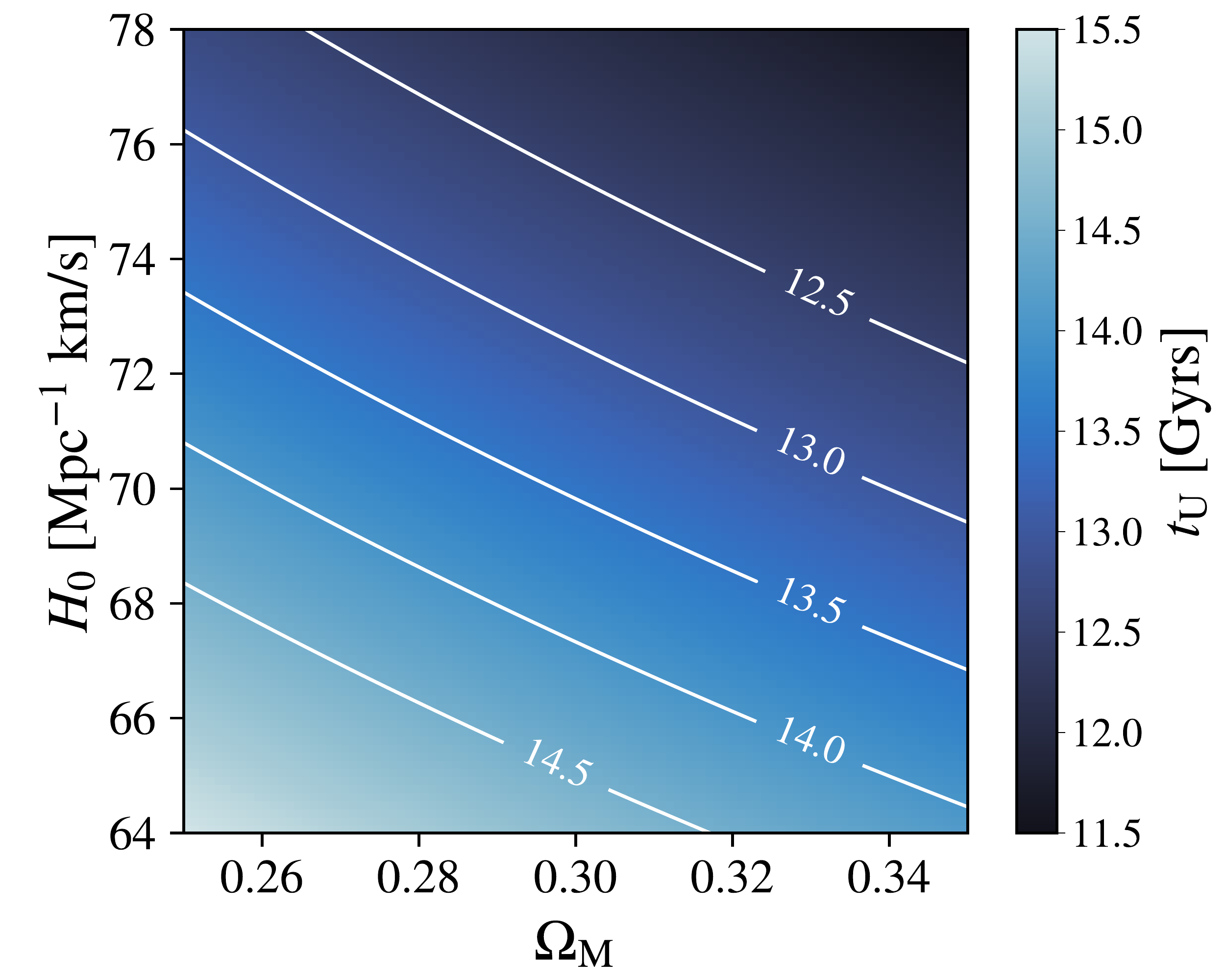}
\includegraphics[width=0.32\textwidth]{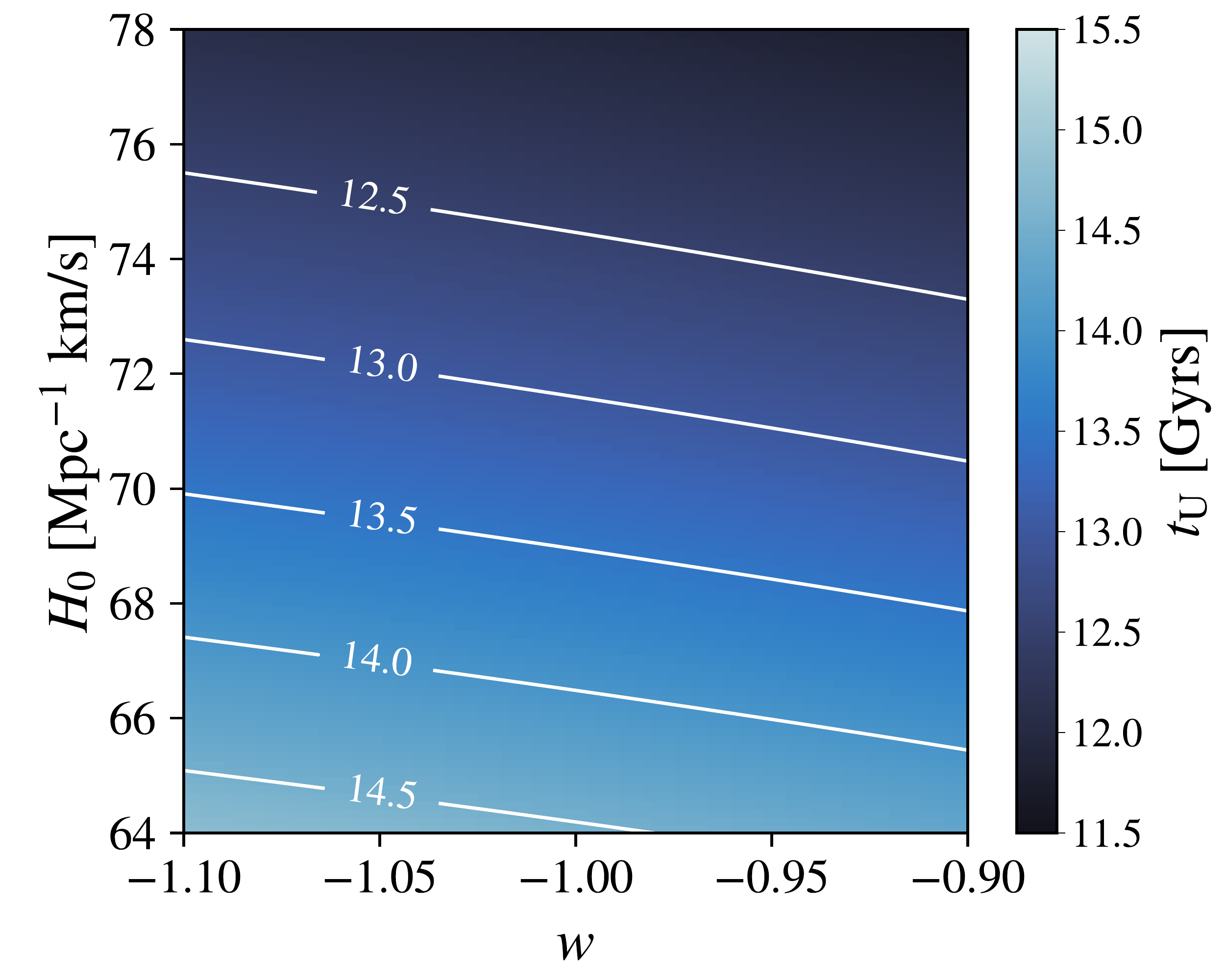}
\includegraphics[width=0.32\textwidth]{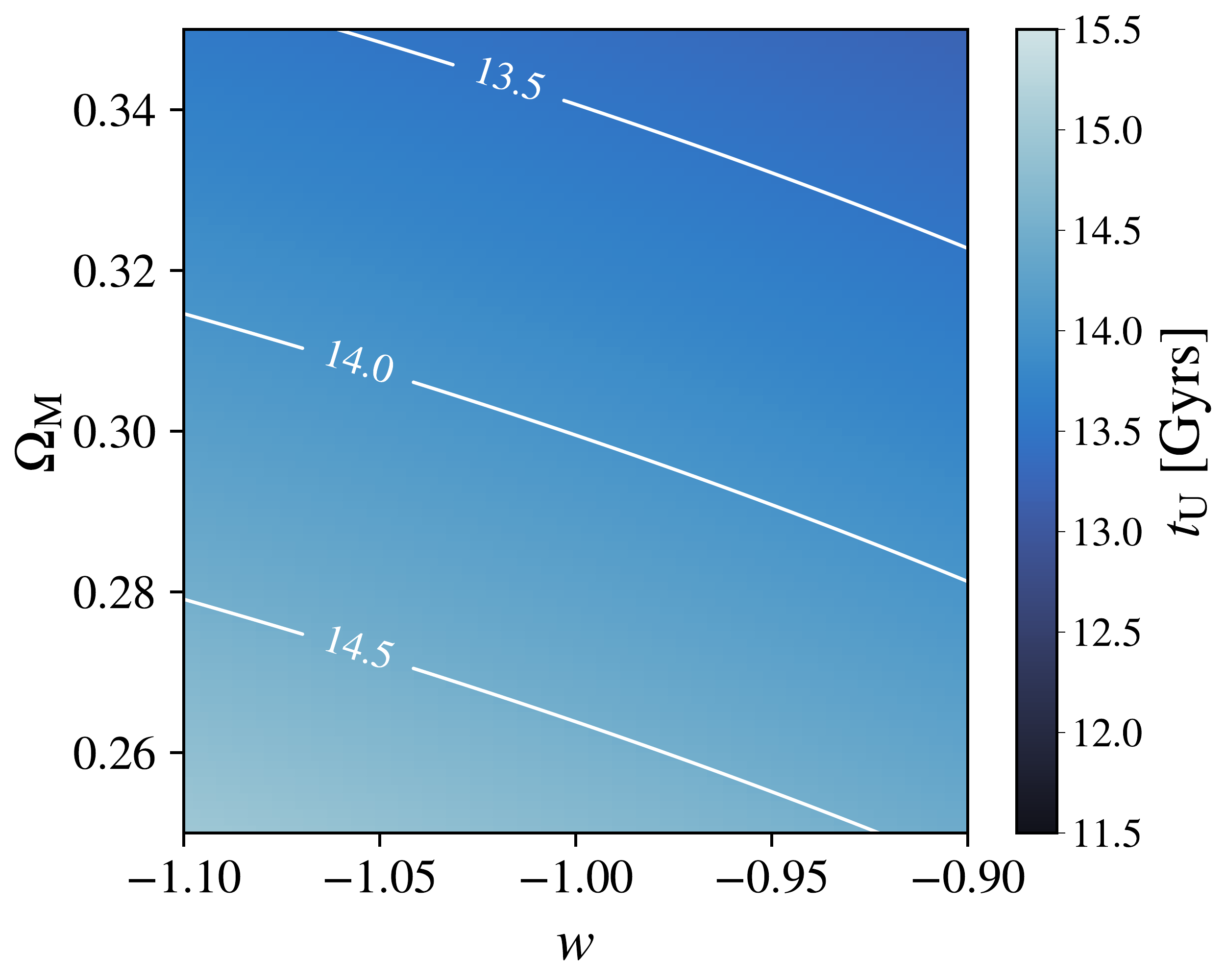}
\caption{Age of the Universe (in Gyr) as function of $H_0$ and $\Omega_{\rm M}$ for $w=-1$ (left panel), $H_0$ and $w$ for $\Omega_M=0.3138$ (central panel), and $\Omega_{\rm M}$ and $w$ for $h=0.6736$ (right panel). When a parameter is not varied, it is fixed to \textit{Planck} $\Lambda$CDM best-fit value. White lines mark contours with constant value of $t_{\rm U}$.}  
\label{fig:tUpars}
\end{centering}
\end{figure*}

 \section{Cosmic Ages and $\boldmath{H_0}$}
\label{sec:tU_tension}
In addition to cosmic distances, the expansion rate of the Universe determines the look-back time. This opens up the possibility to use time (or age) measurements to  weigh in on the  $H_0$ tension. The cosmic chronometers method uses relative ages to determine $H(z)$, but ages can also be used in a complementary way. The look-back time $t$ as function of redshift is given by
\begin{equation}
    t(z) =\frac{977.8}{H_0}\int_0^z \frac{{\rm d}z^\prime }{(1+z^\prime)E(z')}\,{\rm Gyr} ,
    \label{eq:tz}
\end{equation}
with $H(z)$ in ${\rm km\, s^{-1}  Mpc^{-1}}$. Following Eq.~\eqref{eq:tz}, the age of the Universe is $t_{\rm U}\equiv t(\infty)$. We show the dependence of $t_{\rm U}$ on $H_0$, $\Omega_{\rm M}$ and a constant equation of state parameter $w$ for dark energy in a $w$CDM model in Fig.~\ref{fig:tUpars}. It is evident that the strongest dependence is on $H_0$, while $\Omega_{\rm M}$ and $w$ have less influence. 

The integral in Eq.~\eqref{eq:tz} is dominated by contributions from redshifts below few tens, decreasing as $z$ grows. Therefore, any exotic pre-recombination physics does not significantly affect the age of the Universe. On the other hand,  $E(z)$ is bound to be very close to that of a  CMB-calibrated $\Lambda$CDM model at $z\lesssim 2.4$, as shown in the previous section. Hence, a precise and robust determination of $t_{\rm U}$ which does not significantly rely on a cosmological model, in combination with BAO and SNeIa, may weigh in on proposed solutions to the $H_0$ tension. If an independent (and model-agnostic) determination of $t_{\rm U}$  were to coincide with \textit{Planck}'s inferred value assuming $\Lambda$CDM, $\sim 13.8\, {\rm Gyrs}$, alternative models involving exotic physics relevant only in the early Universe would need to invoke additional modifications also of the late-Universe expansion history to reproduce all observations with a high value of $H_0$ as their prediction for $t_{\rm U}$ would be too low.  This is because the value of the integral in Eq.~\eqref{eq:tz} assuming standard physics after recombination cannot be too different from $\Lambda$CDM's prediction once BAO and SNeIa are considered, and then $t_{\rm U}\propto H_0^{-1}$. As we will see below, current measurements of $t_{\rm U}$ are just precise enough to hint at this scenario.

Recently, a value of the age of Universe, $t_{\rm U}=13.5 \pm 0.15 \,(\rm stat.) \pm 0.23 \,(\rm syst.)$  ($\pm 0.27$ when adding statistical and systematic uncertainties in quadrature)  was inferred from a sample of  old globular clusters (GCs) in Refs.~\cite{Valcin_GC, Valcin_GCsyst}.\footnote{This systematic uncertainty was determined using external metallicity spectroscopic measurements of the GCs. We refer the interested reader to Ref.~\cite{Valcin_GCsyst} for more details and an alternative estimate based only on the color-magnitude diagrams of the globular clusters.} This study involves a Bayesian analysis of the properties of 38 GCs, including their age, distance, metallicity, reddening and abundance of $\alpha$-enhanced elements.  $t_{\rm U}$ is inferred from the age of the oldest of these GCs (marginalized over all other parameters and including systematic errors)  estimating  and correcting for  the age of the Universe at the moment of GCs formation,  and   generously  marginalizing over  the small residual dependence on cosmology.

We can confront local $H_0$ measurements with the $t_{\rm U}$ inferred from GCs, since they are related by $H_0t_{\rm U}$, which can be obtained using Eq.~\eqref{eq:tz} and a constraint on $E(z)$ for all the redshifts that contribute significantly to the integral. Redshifts below 2.4 (where  the generic $E(z)$ reconstruction is available) only cover  about 75\% of the age of the Universe.  If we assume that deviations from a $\Lambda$CDM expansion history are driven by the poorly known dark energy component, then  $E(z)$ at $z>2$ is effectively that of an Einstein de Sitter Universe. In this case the  reconstructed $E(z)$ is  perfectly consistent with $\Lambda$CDM and only relatively small deviations are allowed. If we consider more extreme deviations from $\Lambda$CDM, additional data probing the expansion history at higher redshifts would be needed to extend the constraints on the generic $E(z)$ to cover a larger fraction of $t_{\rm U}$.

Hence,  we assume for this study a $\Lambda$CDM expansion rate $E(z)$, using the value  of $\Omega_{\rm M}$ inferred from BAO and SNeIa and its error.\footnote{The  expected  effect of adopting the reconstructed $E(z)$ where available and a $\Lambda$CDM one at higher $z$ is a possible increase of the error-bars on $t_UH_0$ of $\lesssim$ 10\%.}  Note that exotic models modifying only pre-recombination cosmology do not affect directly the late-time $E(z)$ (which remains that of a  $\Lambda$CDM, model) hence our inferred $H_0t_{\rm U}$ also applies to these models.  As an example, we consider early dark energy (EDE) models. In particular, we use the EDE model posterior obtained in Refs.~\cite{Murgia_EDElensing, Smith_EDELSS} for the \textit{Planck} data; the model features three additional cosmological parameters compared to $\Lambda$CDM.
\begin{figure}[!h]
 \begin{centering}
\includegraphics[width=\columnwidth]{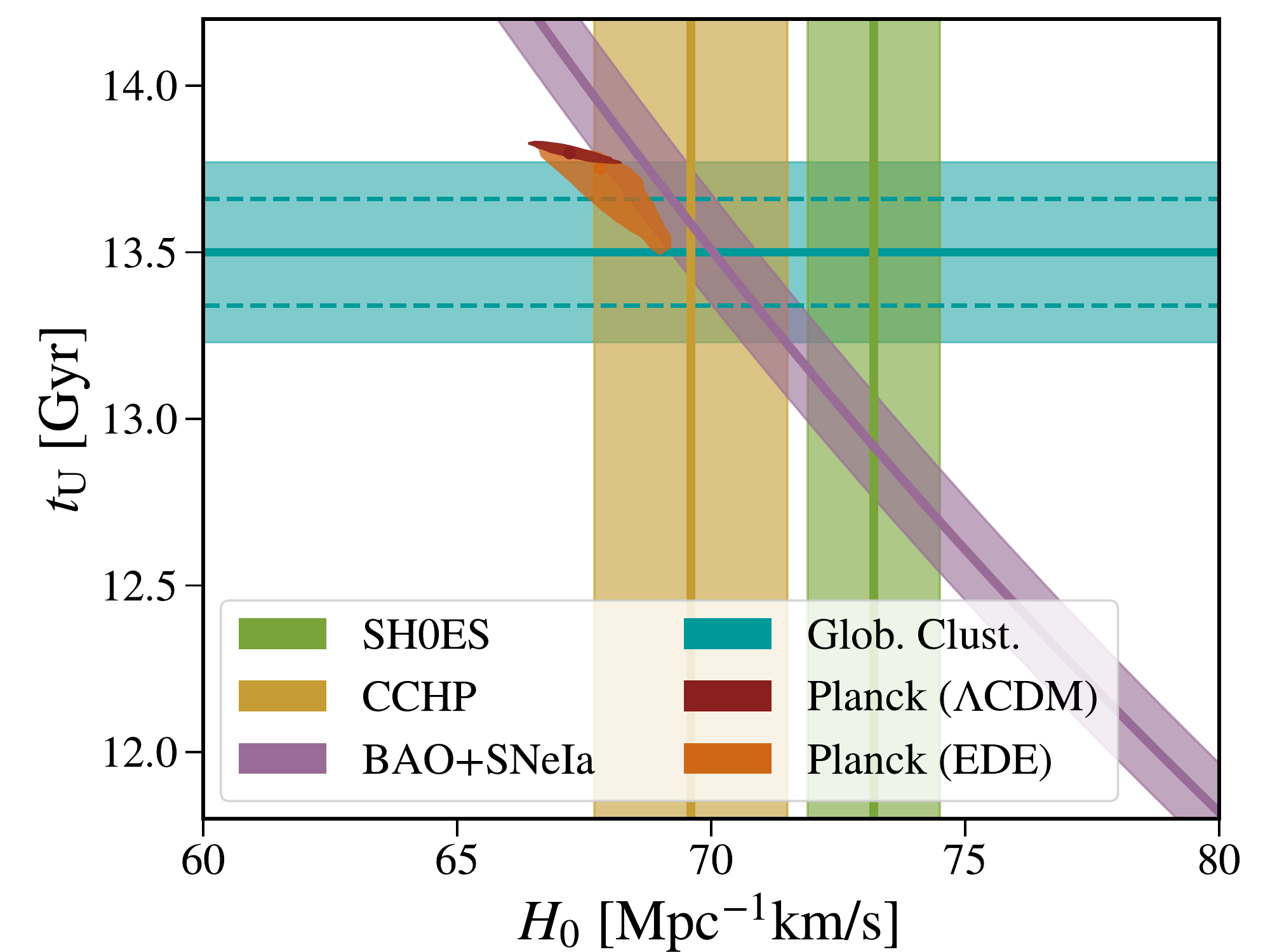}
\caption{68\% confidence level marginalized constraints in the $H_0$-$t_{\rm U}$ plane, from independent measurements, as indicated in the legend. Dashed cyan lines denote the size of the statistical $1\sigma$ errors from globular clusters, while the shaded region also include systematic uncertainties. BAO+SNeIa constraints assume a $\Lambda$CDM cosmology.  We show \textit{Planck} results assuming $\Lambda$CDM (red) and EDE (orange).}  
\label{fig:H0tU}
\end{centering}
\end{figure}

We show 68\% confidence level marginalized constraints on the $H_0$-$t_{\rm U}$ plane from SH0ES, CCHP, GCs, BAO+SNeIa, and \textit{Planck} in Fig.~\ref{fig:H0tU}. We find $H_0t_{\rm U} = 945\pm 11\,{\rm Gyr\,Mpc^{-1}km/s}$ from BAO+SNeIa assuming $\Lambda$CDM, while $H_0t_{\rm U} = 928\pm 7$ and $932\pm 7$ ${\rm Gyr\,Mpc^{-1}km/s}$ from \textit{Planck} assuming $\Lambda$CDM and EDE, respectively. As a reference, combining BAO+SNeIa with SH0ES and TRGB returns $t_{\rm U}=12.93\pm 0.29$ and $t_{\rm U}=13.62\pm 0.42$ Gyr, respectively, while \textit{Planck}'s inferred values are $13.80\pm 0.02$ Gyr ($\Lambda$CDM) and $13.76^{+0.06}_{-0.16}$ Gyr (EDE). 

These results show that for SH0ES to be compatible with BAO+SNeIa the Universe must be significantly younger than inferred by \textit{Planck}, no matter whether $\Lambda$CDM or EDE are assumed; this statement is robust to early-time physics assumptions. The age of the Universe inferred from GCs weakly favors older Universes than SH0ES combined with BAO+SNeIa, but the current systematic error budget is too large to firmly distinguish. There are ongoing efforts to reduce the impact of systematic errors (see e.g.,~\cite{Valcin_GCsyst}), so that GCs constraints on $t_{\rm U}$ have the potential to discriminate among  different scenarios proposed to solve the $H_0$ tension (statistical errors are indicated with dashed lines).

\section{The new cosmic triangles}
\label{sec:triangles}

\begin{figure*}[t]
 \begin{centering}
\includegraphics[width=\columnwidth]{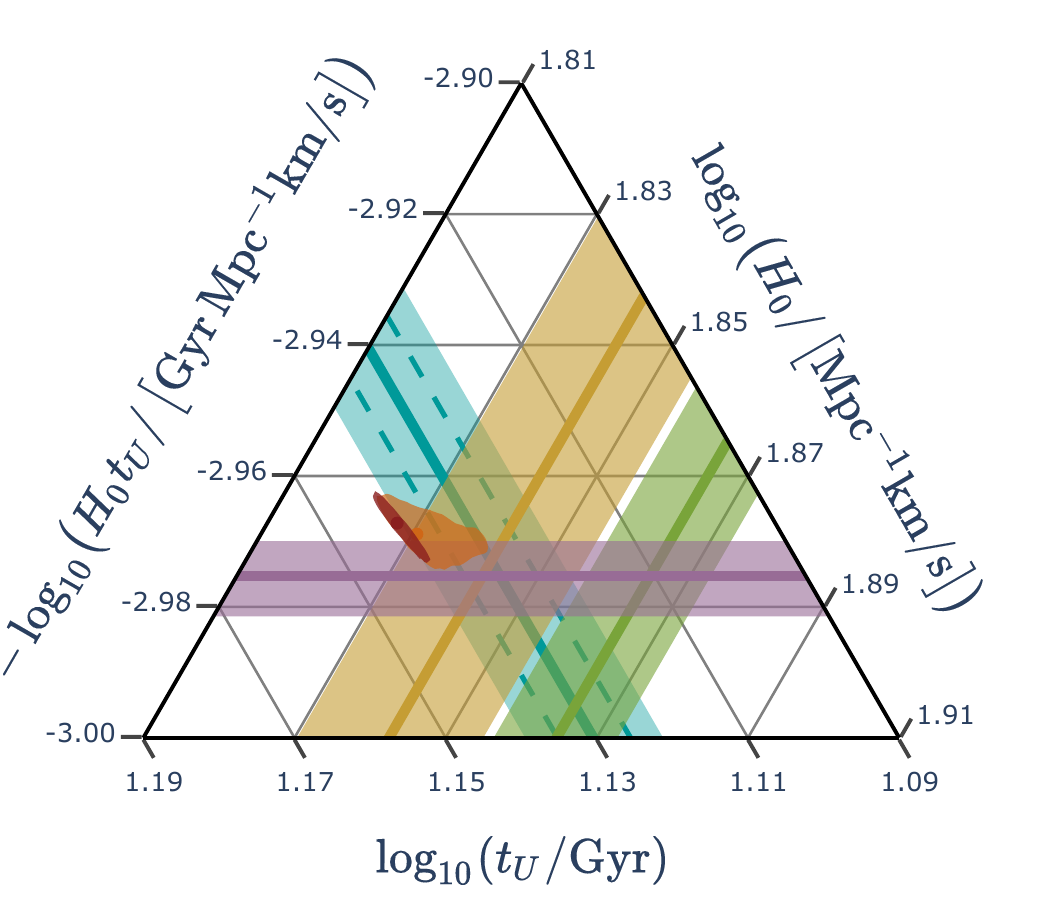}
\includegraphics[width=\columnwidth]{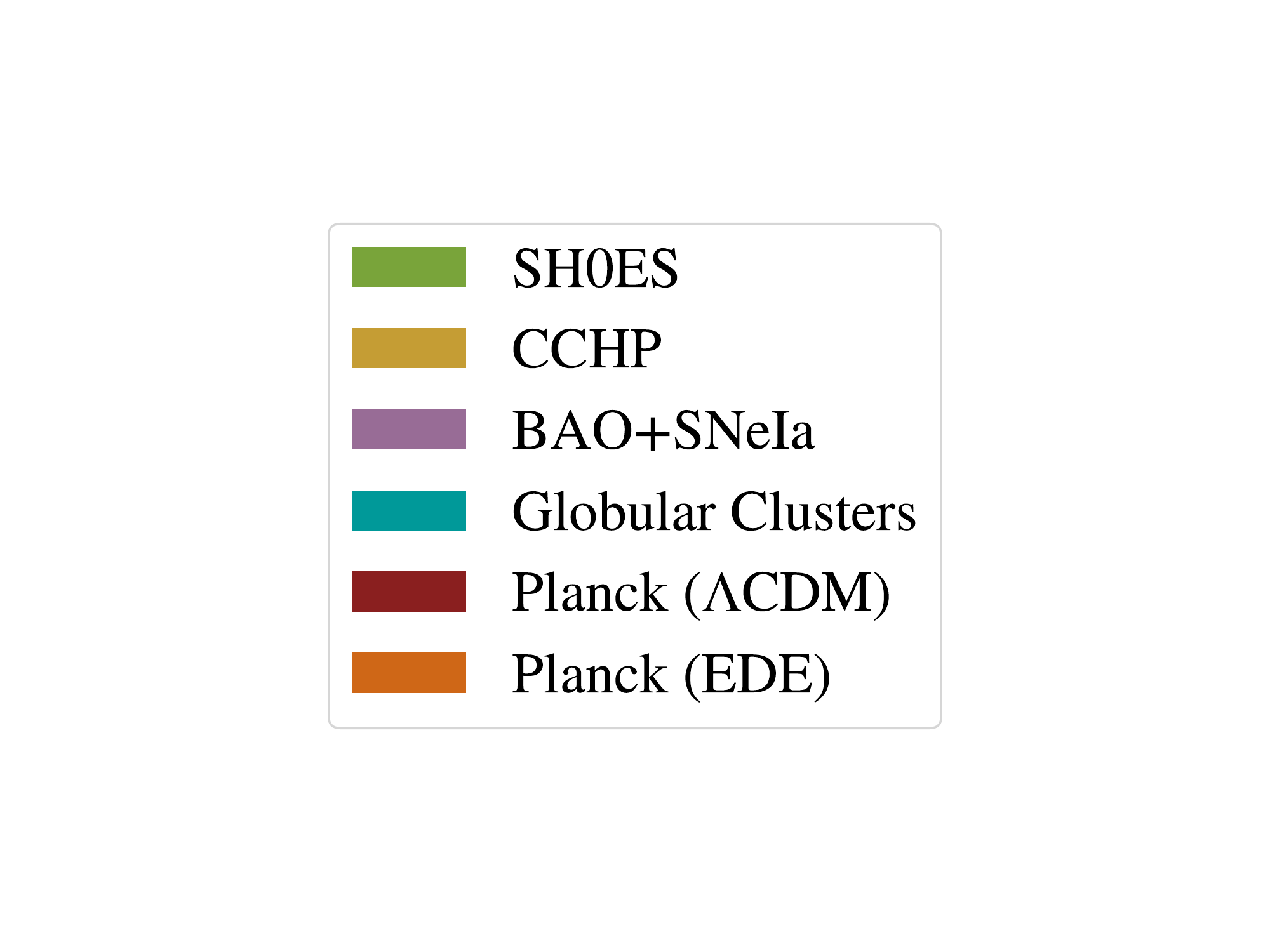}
\includegraphics[width=\columnwidth]{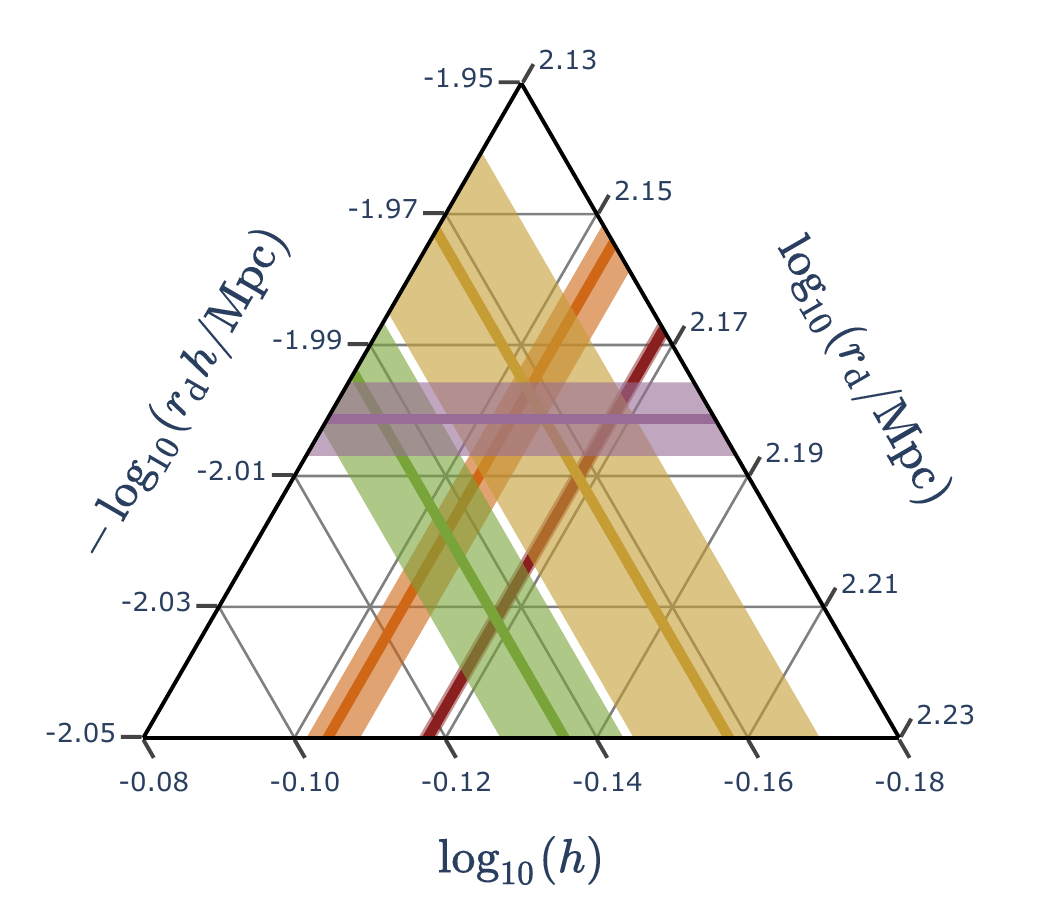}
\includegraphics[width=\columnwidth]{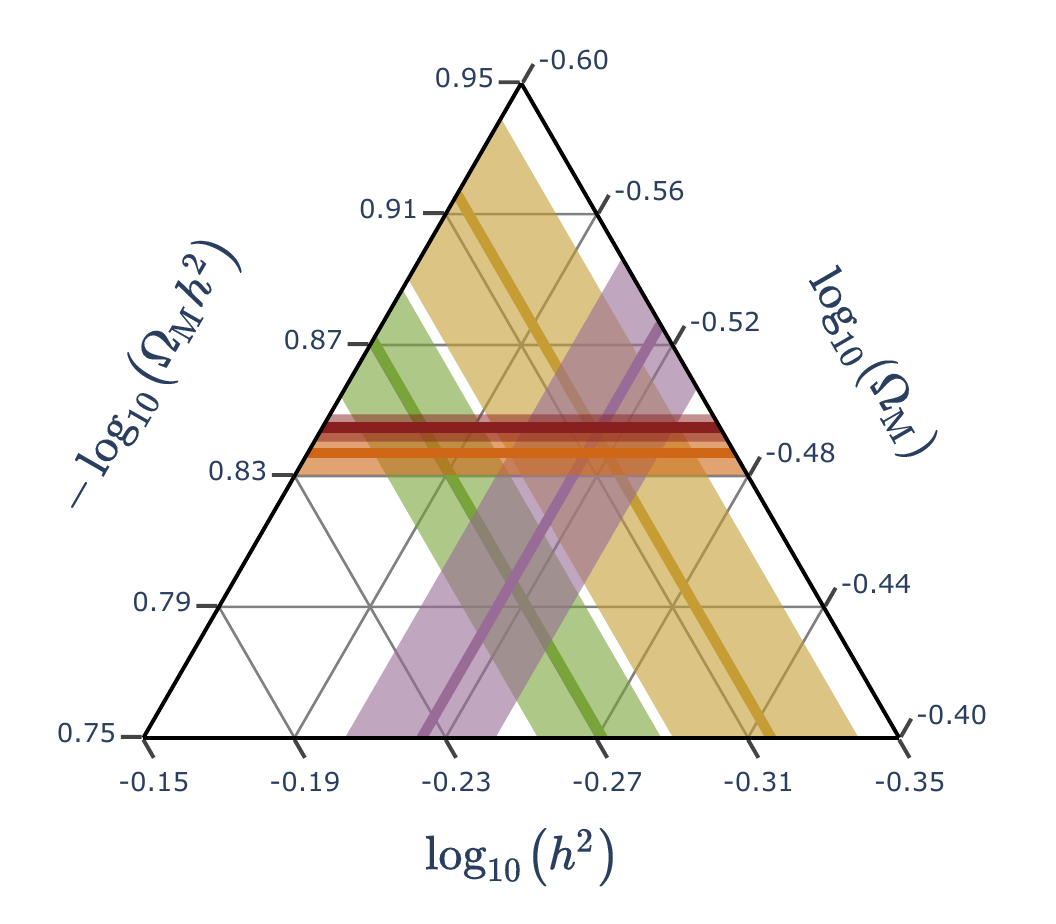}
\caption{68\% confidence level marginalized constraints on the new cosmic triangles: we show the triad corresponding to the age of the Universe and the Hubble constant (upper left), to the sound horizon at radiation drag and the reduced Hubble constant (bottom left), and to the total matter density parameter today and the square of the reduced Hubble constant (bottom right). Note that all points in each figure sum up to 0, while the ticks in the axes determine the direction of equal values for each axis.  }  
\label{fig:triangle}
\end{centering}
\end{figure*}

The $H_0$ tension was reframed as a consistency test between $r_{\rm d}$ (an early-time quantity) and $H_0$ (a late-time quantity), which can be done using a model-agnostic approach, in Ref~\cite{BernalH0}. Similarly, assuming a cosmological model, allows for a similar consistency test between $\Omega_{\rm M}$ and $H_0$ to be performed, as proposed in Ref.~\cite{Lin_H0OmegaM}. With the updated constraints on $E(z)$, $r_{\rm d}h$ and $\Omega_{\rm M}$ obtained in Sec.~\ref{sec:data_and_Ez}, we can revisit these consistency checks. Moreover, the  $H_0$, $t_{\rm U}$ and $H_0t_{\rm U}$  constraints obtained with the $\Omega_{\rm M}$ values inferred from BAO+SNeIa, adds a third consistency test related with $H_0$. 

These three cases are three triads of two cosmological quantities and their product determined independently. These triads are $\left\{t_{\rm U},\, H_0,\, H_0t_{\rm U}\right\}$, $\left\{r_{\rm d},\, h,\, r_{\rm d}h\right\}$, $\left\{\Omega_{\rm M},\, h^2,\, \Omega_{\rm M}h^2\right\}$. Within a given cosmological model (although some of the constraints can be obtained model independently), and in the absence of systematic errors, a generic triad $\left\{a,\, b,\, ab\right\}$ of parameters determined by independent experiments $i$, $j$ and $k$, respectively, is an over-constrained system which must fulfill $a_i\times b_j = (ab)_k$ within statistical uncertainty. This is what makes these triads a powerful  diagnostic tool of  consistency, especially in the context of the $H_0$ tension. Therefore, the cosmological model(s) yielding agreement of all these triads are favored by the data.

Cosmology faced a similar situation in 1999, when information from CMB anisotropies, SNeIa and clusters observations was combined to determine whether the Universe is flat and if  there was evidence for a non-zero cosmological constant~\cite{Bahcall_triangle}. In that case, the triad was $\left\{\Omega_{\rm M},\, \Omega_{\rm k},\, \Omega_\Lambda \right\}$, where $\Omega_\Lambda = 1-\Omega_{\rm M}-\Omega_{\rm k}$ is the density parameter associated to the cosmological constant today.

These triads may be represented in a plane (as done e.g., in Fig.~\ref{fig:H0tU}), but due to the relation between their components, they can be more efficiently represented in a ternary plot. Taking the logarithm of each quantity in the triads of the form $\left\{a,\, b,\, ab\right\}$ (which fulfills $\log_{10}(a)+\log_{10}(b)-\log_{10}(ab)=0$), we can build ternary plots; every point on these ternary plots sum up to 0. This  representation provides an intuitive and illustrative simultaneous look at independent cosmological constraints. We use them to illustrate the state of the $H_0$ tension in each of the three complementary frames that have been discussed. We refer to these ternary plots as the new cosmic triangles. 

Each of the triads discussed in this work involves quantities directly related to $H_0$ and provide different angles to study the $H_0$ tension: in terms of times, distances and the abundance of matter. In interpreting the observational constraints,  we can distinguish between early-time, late-time and local observations, which in turn may  depend on  early-time (pre-recombination),  late-time (low redshift) or  fully local  physics. In all cases, we can use BAO+SNeIa results to link local and early-Universe measurements. Note that the triad corresponding to $h$ and $r_{\rm d}$ is the only one that is agnostic with respect to the choice of a cosmological model for the low-redshift expansion history.\footnote{$r_{\rm d}$ inferred values from \textit{Planck} are largely independent of standard post-recombination physics, as we can see comparing results from standard analyses~\cite{Planck18_pars} with those using only early-Universe information~\cite{StandardQuantities}.}

We show the new cosmic triangles in Fig.~\ref{fig:triangle}; the interpretation of the ternary plots can be eased by comparing this figure with Fig.~\ref{fig:H0tU}. Each side of the triangle corresponds to the logarithm of one of the quantities involved, or their product, and the direction of the ticks in the axes determine the lines of equal value for each quantity. All the constraints shown in these plots (with the exception of the contours corresponding to \textit{Planck} in the upper panel) are bands that refer only to the axis with aligned ticks. The preferred region in the parameter space will be the one with constraints from the three axes overlap. On the other hand, if there is no point in which the constraints referring to all three axes overlap, the measurements are in tension. We can appreciate the tension within $\Lambda$CDM in the triangles corresponding to $r_{\rm d}h$ and $\Omega_{\rm M}h^2$. As expected, considering the region favored by BAO+SNeIa, \textit{Planck} constraints  obtained within  $\Lambda$CDM are consistent with CCHP, but show some tension with SH0ES. The tensions are always smaller in the case of EDE, but not enough for this model to be preferred over $\Lambda$CDM.

Figure~\ref{fig:triangle} clearly shows the synergies of considering the three triads at the same time. The most studied so far has been the one involving $r_{\rm d}$ and $h$, since it was argued that the most promising way to solve the $H_0$ tension was to reduce the value $r_{\rm d}$ while keeping a standard evolution of the low-redshift expansion rate~\cite{BernalH0, Knox_H0hunter}. We can also see that this triangle is the one showing the largest tension between \textit{Planck} assuming $\Lambda$CDM, SH0ES and BAO+SNeIa, and the one for which models like EDE  show promise. The triangle including $\Omega_{\rm M}$ shows a smaller tension: combining BAO+SNeIa with SH0ES (CCHP) we find $\Omega_{\rm M}=0.159\pm 0.009$ ($\Omega_{\rm M}=0.144\pm 0.01$), which is in 1.8$\sigma$ (0.1$\sigma$) tension with \textit{Planck}'s constraint assuming $\Lambda$CDM. The tension reduces to 1.5$\sigma$ when compared to the \textit{Planck} results assuming EDE. Since BAO+SNeIa constrain $E(z)$ at low redshift to be very similar to (and fully consistent with) the best fit of \textit{Planck} assuming $\Lambda$CDM,  this tension is fully sourced by the $H_0$ tension, no matter the cosmological model under consideration.

However, the situation for the triad involving the age of the Universe is different. As argued above, modifications of the early-Universe cosmology do not directly change the age of the Universe. This is why \textit{Planck} EDE posteriors overlap with those assuming $\Lambda$CDM (extending along the direction of constant $\Omega_{\rm M}$, i.e., the constraint on $H_0t_{\rm U}$ from BAO+SNeIa). In this representation, the region of  overlap of \textit{Planck}, BAO+SNeIa and GCs posteriors is in large tension with SH0ES. However, current determinations of $t_{\rm U}$ alone are not precise enough to definitively disfavor the combination of SH0ES with BAO+SNeIa. 

Finally, Fig.~\ref{fig:triangle} clearly indicate that  if GCs were to  still return a high value of $t_{\rm U}$ but with reduced  error-bars, deviations from $\Lambda$CDM  that only affect pre-recombination physics will not be enough to reconcile all the measurements. If this  will turn out to be the case, a combination of both, high and low redshift modifications to the  $\Lambda$CDM model may be required to solve the $H_0$ tension.
Alternatively one would have to look into much more local effects, such as those affecting the distance ladder calibration and in particular effects or processes which may be responsible for the mis-match between CCHP and SH$0$ES.

\section{Conclusions}
\label{sec:conclusions}
The discrepancies between model-independent measurements and  model-dependent inferred values of $H_0$ from different experiments (each of them sensitive to different physics and systematic errors) might be a hint for the need of modifying the standard $\Lambda$CDM model. The most promising deviations from $\Lambda$CDM proposed to solve such tensions involve a boost in the expansion rate before recombination, as to lower the value of $r_{\rm d}$ and reconcile the direct and the inverse distance ladder. However, we argue in this work,  there is a  more varied phenomenology, that goes well beyond $r_d$,  to be matched by any new physics put forward to solve the $H_0$ tension, especially regarding cosmic ages: the trouble goes beyond $H_0$.

We update agnostic reconstructions of the evolution of the expansion rate of the late-time Universe with recent BAO and SNeIa measurements, extending the reconstruction up to $z\sim 2.4$. We find that BAO and SNeIa constrain the evolution of $H(z)$ to be fully consistent  with  the one from $\Lambda$CDM \textit{Planck}'s best-fit prediction:  any possible deviation must be well below the 5\%(10\%) level at $z<0.8$ ($z<2.4$). This further supports previous claims that  modifications of the expansion rate at low redshifts are disfavored by the data (see e.g.,~\cite{BernalH0, Poulin_H0, Knox_H0hunter}). In the coming years, line-intensity mapping~\cite{Karkare_IMBAO, Munoz_vao, Bernal_IM, Bernal_IM_letter, Silva_LIM_whitepaper}, quasar observations~\cite{Risaliti_QSOH0, Risaliti_QSOH0nature} and strong lensing systems~\cite{Chen_SlensEz} will probe significantly higher redshifts, allowing for agnostic analyses like this one to be extended up to $z\sim 10-20$ (covering effectively $>90 $\% of the Universe's history). 

We discuss the impact of a recent, almost cosmology-independent, inference of the age of the Universe from the age of the oldest globular clusters. While the relation between $H_0$ and $r_{\rm d}$ can be addressed with modifications of the early-time physics, $t_{\rm U}$ is dominated by the expansion rate at $z\lesssim 30$, hence insensitive to high-redshift cosmology. The  $t_{\rm U}$ determination is also insensitive to effects such as cosmological dimming (e.g., violations of the Etherington relation), cosmological screening,  deviations from general relativity at large scales affecting growth of structures and  any phenomenology affecting cosmological distance measures.   Therefore, if a high $t_{\rm U}$ were to be measured reliably and with small enough error-bars, it would disfavor models with high $H_0$ and standard low-redshift physics. In this case then  both, pre- and post-recombination modifications to $\Lambda$CDM, may  be required to reconcile all measurements.  Alternatively  one  would  have  to  invoke much  more local  effects (be these cosmological, see  e.g.,~\cite{Desmond_H0screening,Desmond_H0screening_e1, Desmond_H0screening_e2, Desmond_screening_trgb}, or astrophysical, in particular effects or processes which may be responsible for the mis-match between CCHP and SH$0$ES) affecting the local $H_0$ determination only, while leaving all other cosmological observations unchanged. 

In such case, viable solutions to the $H_0$ trouble  will fall in either of  two classes of very different nature: local and global. Global solutions,  would have to invoke new physics beyond $\Lambda$CDM which affect the entire Universe history from before recombination all the way to the low-redshift, late-time Universe. Modifying only early-time physics will not be enough. Because of their global nature, such solutions affect quantities  well beyond $H_0$, but would be highly constrained by the wealth of high-precision cosmological observations available.
Local solutions on the other hand, leave unaffected the global properties of cosmology; as such either do not require new physics beyond $\Lambda$CDM (and thus fall in the realm of astrophysics),  or include new physics which only affect very local observations.

A program to improve the inference of $t_{\rm U}$ and reduce the systematic uncertainties, may give this measurement enough power to  discriminate between these two different kinds of viable solutions for the $H_0$ tension.

Finally we identify three triads of independently-measured quantities, relating $H_0$ with $t_{\rm U}$, $r_{\rm d}$, $\Omega_{\rm M}$, respectively. Each of these triads is an over-constrained system, hence we propose the use of ternary figures (the new cosmic triangles) to  report and visualize the constraints. These new cosmic triangles allow for a simultaneous and easy-to-interpret visual representation of constraints on different  yet related quantities. We hope that  this representation will help to guide   further efforts to find a solution to  the trouble of  (and beyond) $H_0$.

\acknowledgments
The authors thank Tristan L. Smith, Vivian Poulin and Geoff C.-F. Chen for comments on last versions of this manuscript. JLB is supported by the Allan C. and Dorothy H. Davis Fellowship. This work is supported in part by  MINECO grant PGC2018-098866-B-I00 FEDER, UE. LV acknowledges support by European Union's Horizon 2020 research and innovation program ERC (BePreSySe, grant agreement 725327). ICC researchers acknowledge ``Center of Excellence Maria de Maeztu 2020-2023” award to the ICCUB (CEX2019-000918-M). This work was supported at Johns Hopkins by NSF Grant No.\ 1818899 and the Simons Foundation. The work of BDW is supported by the Labex ILP (reference ANR-10-LABX-63) part of the Idex SUPER, received financial state aid managed by the Agence Nationale de la Recherche, as part of the programme Investissements d’avenir under the reference ANR-11-IDEX-0004-02; and by the ANR BIG4 project, grant ANR-16-CE23-0002 of the French Agence Nationale de la Recherche. The Center for Computational Astrophysics is supported by the Simons Foundation.

\bibliography{Refs.bib}
\bibliographystyle{utcaps}
\bigskip

\end{document}